\documentclass[aps,prl,twocolumn,showpacs,groupedaddress]{revtex4-1}
\usepackage{graphicx}
\usepackage{epstopdf}
\usepackage{color}
\usepackage{url}
\usepackage{hyperref}
\usepackage[T1]{fontenc}
\usepackage[latin1]{inputenc}
\usepackage{float}

\newcommand{\abs}[1]{\left\vert #1 \right\vert}

\def\figref{Fig.~}
\def\eqnref{Eq.~}

\usepackage{amsmath}	
\begin{document}
\title{Optimised multi-ion cavity coupling}

\author{S. Begley}
\author{M. Vogt}
\author{G. K Gulati}
\altaffiliation{1984.gurpreet@gmail.com}
\author{H. Takahashi}
\author{M. Keller}

\begin{abstract}
  Recent technological advances in cavity quantum electrodynamics (CQED) are paving the way to utilise multiple quantum emitters confined in a single optical cavity.
In such systems it is crucially important to control the quantum mechanical coupling of individual emitters to the cavity mode. In this regard, combining ion trap technologies with CQED provides a particularly promising approach due to the well-established motional control over trapped ions. 
Here we experimentally demonstrate coupling of up to five trapped ions in a string to a high-finesse optical cavity.  By changing the axial position and spacing of the ions in a fully deterministic manner, we systematically characterise their coupling to the cavity mode through visibility measurements of the cavity emission. In good agreement with the theoretical model, the results demonstrate that the geometrical configuration of multiple trapped ions can be manipulated to obtain optimal cavity coupling. Our system presents a new ground to explore CQED with multiple quantum emitters, enabled by the highly controllable collective light-matter interaction.

\end{abstract}


\maketitle

Over the past decade, techniques to localise quantum emitters inside high-finesse optical cavities have been extensively explored in the realm of cavity quantum electrodynamics (CQED) \cite{Miller2005,Keller2004,Yoshie2004,Colombe2007,Faraon2012}. The CQED systems constitute one of the simplest forms of light-matter interaction, yet provide highly versatile platforms for studies in fundamental physics \cite{Raimond2001} and applications in quantum information \cite{Kimble2008}. The latter is highlighted by the recent demonstrations of quantum interfaces and elementary quantum networks with atomic systems \cite{Wilk2007,Ritter2012,Stute2013}.

So far there have been two major strategies in atomic CQED: either employing a single atom or large ensemble of atoms. In the latter case only the averaged contribution from many atoms is of interest and individual atoms are not controlled \cite{Herskind2009,McConnell2015}.  Even though the idea to couple multiple atoms deterministically to a single optical mode dates back to the well-known Dicke model of superradiance \cite{Dicke1954}, experimental efforts have started only recently \cite{Casabone2013,Reimann2015,Casabone2015}. In general, many-body interactions mediated by cavity photons in multi-atom CQED give rise to far richer physics than single-atom CQED (e.g. \cite{Morrison2008,Gopalakrishnan2011}) and find many applications, such as quantum logic gates \cite{Pellizzari1995,Zheng2004}, nonclassical light sources \cite{Fernandez-Vidal2007,Habibian2011} and entanglement generation schemes \cite{Gonzalez-Tudela2013,DallaTorre2013}. Furthermore, this system fits naturally into the quantum network architecture where each nodal station should possess a multi-qubit quantum register interfaced by an optical cavity \cite{Kimble2008,VanEnk1997}.
\begin{figure}[t!]
 \includegraphics[width=\columnwidth]{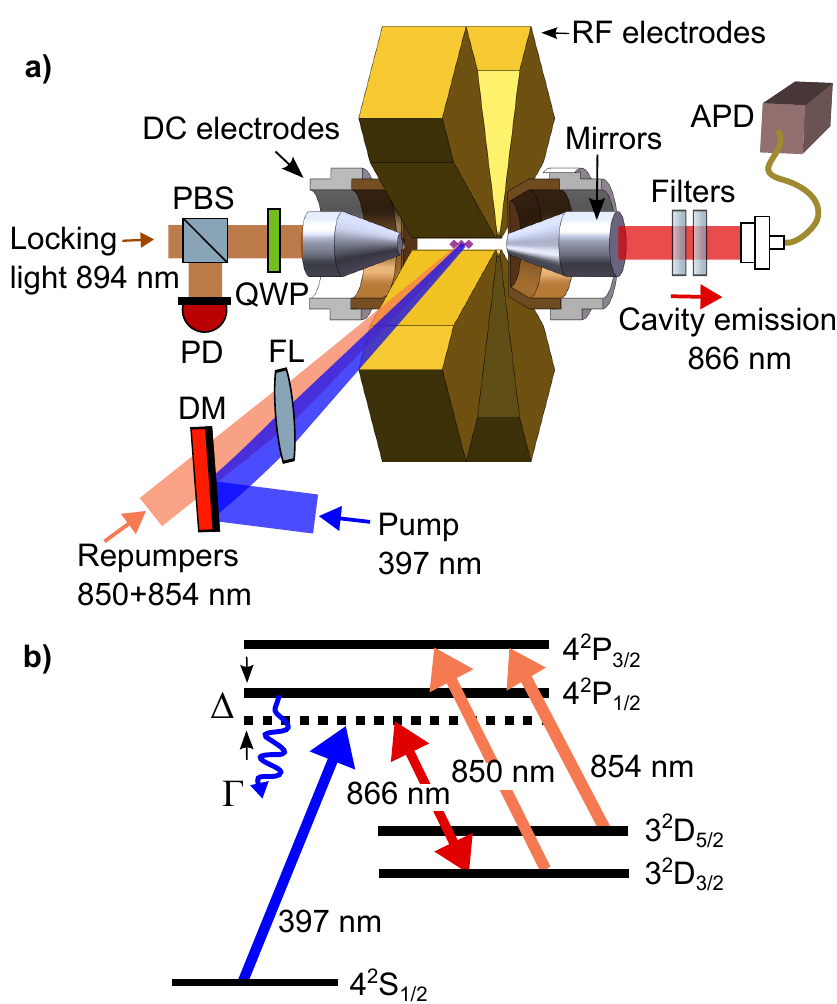}
 \caption{\label{fig:setup} a) Experimental setup. Pump and repump laser beams are overlapped and sent oblique to the trap axis. The cavity emission at 866\,nm is detected with an avalanche photodiode (APD). A reference laser at 894\,nm is used for locking the cavity frequency. A combination of filters are used to remove background light from the cavity emission. DM: dichroic mirror, FL: focusing lens, PBS: polarization beam splitter, PD: photodiode, QWP: quarter wave plate. b) Level scheme of $^{40}\text{Ca}^{+}$ indicating wavelengths relevant for the experiment.}
\end{figure}
The major challenges in multi-atom CQED lie in controlling couplings of the individual atoms to the cavity mode in its standing wave pattern while minimizing unwanted excess motion. In that regard, ion trap systems have a particular advantage in their unparalleled motional control over trapped ions. A linear string of single ions can be easily localized in the Lamb-Dicke regime i.e. the extent of the motional wavefunctions of the trapped ions is smaller than the relevant optical wavelength \cite{Leibfried2003}.
Despite the fact that the inter-ion distances cannot be made uniform in general, for a moderate number of ions, the overall confinement can be optimised in order to couple all the ions with nearly maximal strength to the cavity mode. 

In this article we present an experiment to deterministically control the position and spacing of single ions trapped inside a high finesse optical cavity. In good agreement with the theory, these results demonstrate that we can achieve a simultaneous coupling of multiple ions to the cavity mode with a high degree of controllability.  We also show that the presence of multiple ions can significantly enhance their localization due to their mutual Coulomb interaction. 

\section{Results}
\subsection{Experimental setup.}

\figref\ref{fig:setup}a shows our experimental setup. A string of ${}^{40}\text{Ca}^{+}$ ions are trapped in a linear Paul trap with its axis aligned to coincide with the axis of an optical cavity.  This configuration enables multiple ions to be coupled to the cavity mode simultaneously.
In our ion trap, the radial confinement is provided by four blade-shaped
electrodes with their pointed edges arranged on a square centered around the trapping region. The distance from the trap center to the rf electrodes is 475\,$\mu$m. A pair of dc endcap electrodes with a separation of 5\,mm provides confinement in the axial direction.
The radial secular frequency is 1.23\,MHz whereas the axial secular frequency is changed over 400--620\,kHz during the experiment in order to vary the axial spacing of the ions. The stray electric fields in the trap 
that result in the ions' micromotion 
are compensated by applying correctional dc voltages onto the rf electrodes.
The cavity mirrors are enclosed inside the hollow inner structure of the dc endcap electrodes (see \figref~\ref{fig:setup}a). 1 mm diameter openings at the centers of the endcaps allow a line of sight between the mirrors which form an optical cavity symmetrically around the trapped ions with a cavity length of 5.3 mm. Incorporating the mirrors in this way protects the trapping region from potential distortion by the dielectric surfaces of the mirrors \cite{harlander2010trapped}. 
The two cavity mirrors have transmissivities of 100 ppm and 5 ppm at 866 nm, leading to a cavity finesse of $\sim$60,000 and a linewidth of 470\,kHz.

\figref\ref{fig:setup}b shows the level scheme and relevant transitions in $^{40}\text{Ca}^{+}$. 
The ground state $\text{S}_{1/2}$, the excited state $\text{P}_{1/2}$ and the metastable state $\text{D}_{3/2}$ form a three-level $\lambda$\,-\,system.
Here the pump beam at 397\,nm on $\text{S}_{1/2}$-$\text{P}_{1/2}$ serves two purposes: Doppler cooling of the trapped ions and driving one arm of the cavity-assisted Raman transition.  
The former is accomplished by adjusting the detuning $\Delta$ of the pump beam to $\sim\Gamma/2$ on the red-side of the transition. Here $\Gamma$(=$2\pi\times 22.3$ MHz) is the total decay rate of the $\text{P}_{1/2}$ state. An additional laser beam with far detuning ($6.5\,\Gamma$) is used to stabilize the string of multiple ions for the measurements shown in \figref\ref{fig:multi-ion-coupling} and \figref\ref{fig:3-4-5-coupling}.
The frequency of the pump laser is stabilized through a scanning cavity lock \cite{Seymour2010} to a reference laser at 894 nm which in turn is locked to a cesium atomic transition.
A branch of the same reference laser is used to stabilize the cavity resonance frequency with respect to the $\text{P}_{1/2}$-$\text{D}_{3/2}$ transition. By satisfying the Raman condition, i.e. the pump and cavity detunings being equal, a population in $\text{S}_{1/2}$ can be coherently transferred to $\text{D}_{3/2}$.  
This results in the emission of a 866\,nm photon into the cavity mode \cite{Keller2005}.
 A combination of 850\,nm and 854\,nm repumper lasers is employed to depopulate the metastable $\text{D}_{3/2}$ state (lifetime\,$\approx$\,1\,s) to provide continuous cooling and cavity emission. 
Due to the asymmetric transmissivities of the cavity mirrors, the photon emission in the cavity is predominantly outcoupled to one direction. This directed continuous photon stream is further coupled to a single mode fibre and its rate is measured by an avalanche photodiode (APD).
\begin{figure}[t]
\begin{center}
\includegraphics[width=\columnwidth]{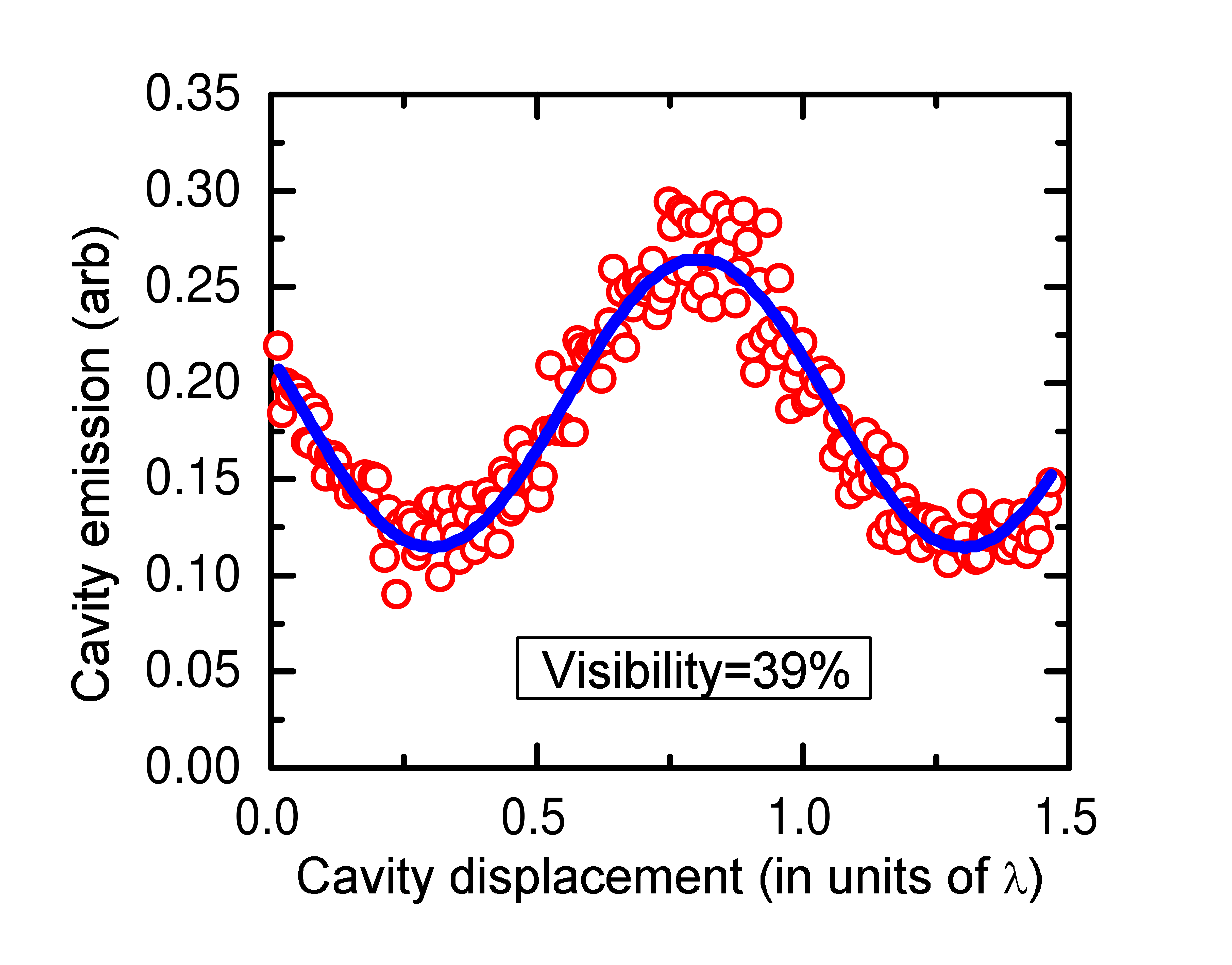}
\caption{\label{fig:singleion}Cavity emission rate $R_{em}$ as the cavity
is translated relative to the position of a single ion ($\lambda$\,=\,866\,nm). The axial secular frequency is 620\,kHz. A visibility of 39\% is obtained from a sinusoidal fit. This corresponds to a temperature of $\approx$\,1.2\,$T_{\text{D}}$ where $T_{\text{D}}$ is the Doppler temperature. Note that the far-detuned cooling beam is not used in this measurement.}
\end{center}
\end{figure}

\subsection{Coupling a single ion to the cavity.}

The coupling rate $g$ between a single ion and the cavity field varies
with the local field strength at the position of the ion \cite{Guthoerlein2001}. It is
characterised by a standing wave pattern along the cavity
axis ($z$).
\begin{equation}
g(z) = g_{0}\cos\left(kz\right), \label{eq:gz}
\end{equation}
where $k\,=\,2\pi/\lambda$ is the wave number of the cavity field and $2g_0$ is the vacuum Rabi frequency. With our cavity geometry, $g_0$ is $2\pi\times 0.9$ MHz.
The variation of $g$ along radial direction is neglected as the cavity field varies axially on a much shorter length scale than radially. 
The spatial wavefunction of a single ion $\Psi(z)$ in a harmonic potential is described by a Gaussian distribution with a variance of
\begin{equation}
(\Delta z)^2 \approx \frac{2k_{B}T}{m\omega_{sec}^2},\label{eq:eignf}
\end{equation} 
where $\omega_{sec}$ is the axial secular frequency and $T$ is the temperature of the ion. Due to this spatial fluctuation of the ion, the average cavity emission rate $R_{em}$ is
\begin{figure}[t]
\includegraphics[width=\columnwidth]{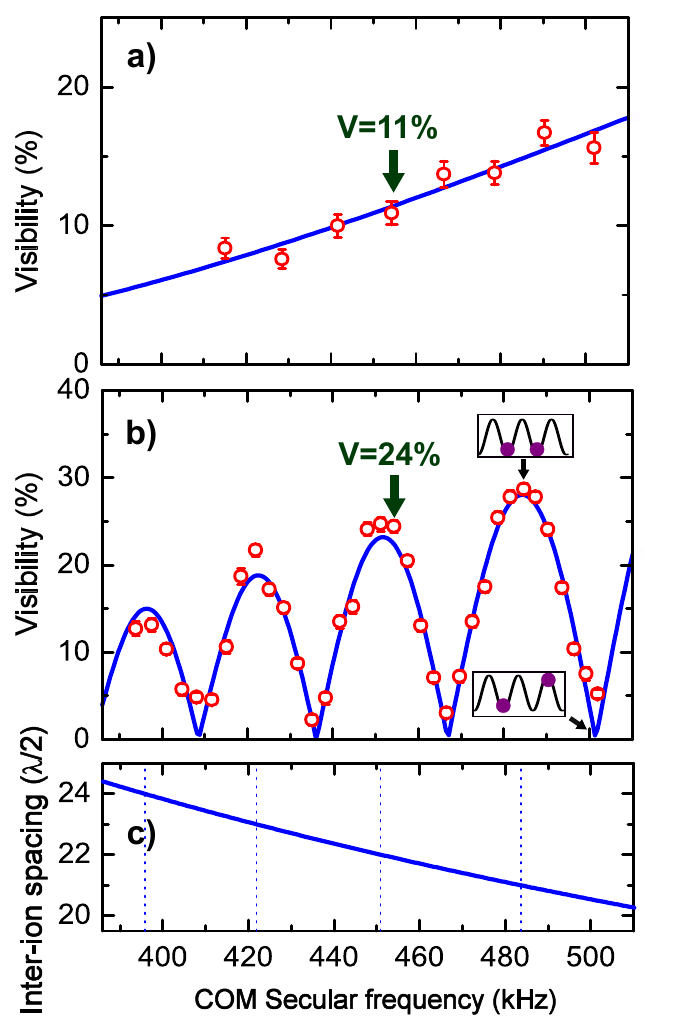}
 \caption{\label{fig:multi-ion-coupling} Visibility measurements for a) a single ion and b) two ions as a function of the COM secular frequency. Error bars are derived from the fitting errors. The blue curves are fits based on our theoretical model (See Methods). The temperatures extracted from the fits are $\sim 1.5\,T_{\text{D}}$ in both cases. The increase in temperature compared to \figref~\ref{fig:singleion} is due to the presence of the far detuned laser.
 The inset figures in b) illustrate the relationships between the ions' positions and the spatial variation of $g^{2}(z)$ at a local maximum and minimum of the visibility.
 c) The inter-ion spacing between two ions. The dotted vertical lines indicate the COM secular frequencies at which the spacing becomes an integer multiple of $\lambda/2$.}
\end{figure}
 proportional to the convolution of $g(z)^2$ and $\abs{\Psi(z)}^2$: 
%
\begin{equation}
 R_{em}(z) \propto g_{0}^{2}\left(1+e^{-k^{2}(\Delta z)^{2}}\cos(2kz)\right). \label{eq:cavityemission}  
\end{equation}
$R_{em}(z)$ is maximal (minimal) when the ion is located at an antinode (node) of the cavity field for a given $\Delta z$.

As shown in \figref~\ref{fig:singleion}, this periodic dependence of $R_{em}(z)$ on $z$ is observed at the APD count, as we translate the entire cavity around a single ion while keeping the absolute position of the ion and the cavity length both constant. 
The contrast of this pattern reveals how well the ion is localized and is quantified by the visibility  $V = \frac{R_{em}^{max}-R_{em}^{min}}{R_{em}^{max}+R_{em}^{min}}$ where $R_{em}^{max}$ and $R_{em}^{min}$ are the maximum and minimum of the observed count rate. For a single ion \cite{Guthoerlein2001}, 
\begin{equation}
 V = e^{-k^{2}(\Delta z)^{2}} \; .\label{eq:visibility}
\end{equation}
The visibility value of 39\% shown in \figref~\ref{fig:singleion} corresponds to $\Delta z$=133 nm and a temperature of $\approx$\,1.2\,$T_{\text{D}}$ where $T_{\text{D}}$ is the Doppler temperature (= 535 $\mu$K), confirming good localization of a single ion inside the cavity.
The visibility will be used as a figure of merit throughout the following analysis.

\subsection{Coupling multiple ions to the cavity.}

When multiple ions are weakly coupled to the cavity mode (i.e. $g_0 < \Gamma$) and repumped continuously, the total cavity emission rate is a sum in the form of $\sum_{i} R_{em}(z_i)$ where each term $R_{em}(z_i)$ represents the emission rate of each individual ion (See \eqnref(\ref{eq:cavityemission})).
The visibility for multiple ions is no longer a simple function of the localization $\Delta z$ as in \eqnref(\ref{eq:visibility}). Instead the inter-ion spacing with respect to the cavity wavelength plays an important role.

\begin{figure}
 \includegraphics[width=\columnwidth]{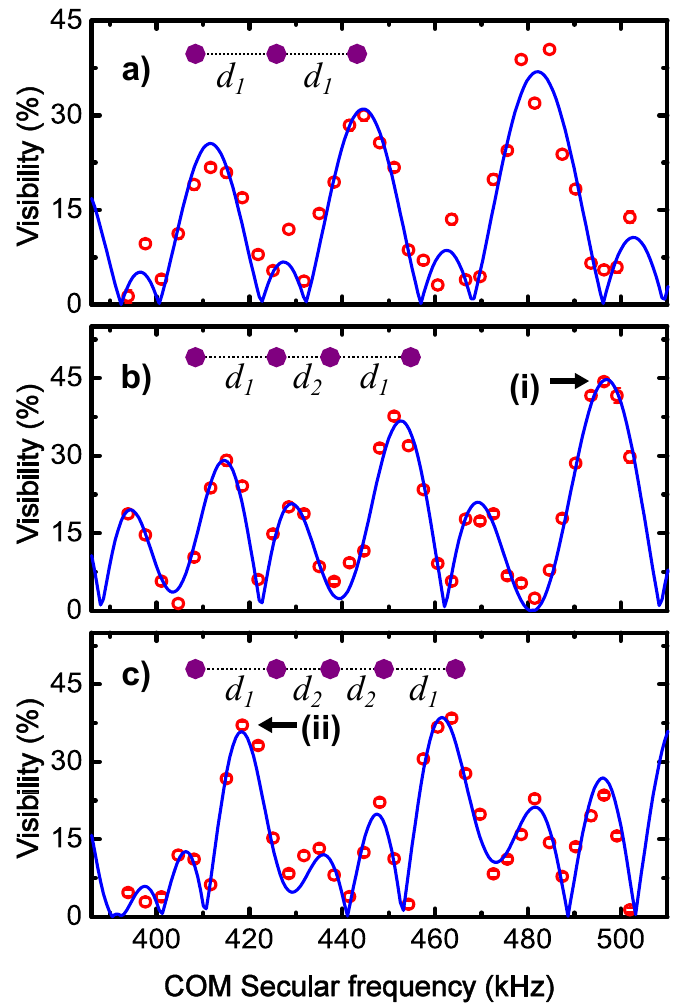}
 \caption{\label{fig:3-4-5-coupling} Visibility measurements for a) three, b) four and c) five ion strings. The fits assume that the temperatures of all normal modes are identical. The extracted temperatures are a) 1.56 $T_{\text{D}}$, b) 1.57 $T_{\text{D}}$, c) 1.72 $T_{\text{D}}$, showing a slight increase with the number of ions. The insets illustrate the configurations of ion strings. In particular with four and five ions the inter-ion spacing is not uniform, i.e. $d_1 \neq d_2$.
    }
\end{figure}

\figref\ref{fig:multi-ion-coupling}a and b show the visibility measurements for one and two ions as the secular frequency of the center of mass (COM) mode is varied. For each data point shown as a red circle, we performed a visibility measurement with a fixed COM secular frequency in the same manner as the one shown in \figref~\ref{fig:singleion} and extracted the visibility value from a sinusoidal fit. Note that the range of secular frequencies is chosen in accordance with the measurements in \figref~\ref{fig:3-4-5-coupling} such that a string of up to five ions does not becomes kinked at high COM secular frequencies. \figref\ref{fig:multi-ion-coupling}c shows the calculated inter-ion distance for two ions as a function of COM secular frequency. The local maxima of the visibility in \figref~\ref{fig:multi-ion-coupling}b are attained when the ions are spaced by an integer multiple of $\lambda/2$, such that they can sit simultaneously at nodes or antinodes of the cavity field. At these inter-ion spacings, the two ions are coupled to the cavity mode with equal strength. 
In contrast, when the ions are separated by a half-integer multiple of $\lambda/2$, the total emission rate from the two ions remain constant regardless of the cavity displacement and as a result the visibility vanishes. 

We have also observed greater values of visibility for two ions at the local maxima when compared to the measurements with a single ion at the same COM secular frequencies. For example, at 454 kHz in \figref~\ref{fig:multi-ion-coupling}a and b, the visibility value is 24\% for two ions whereas it is only 11\% for a single ion.   
This shows that the mutual Coulomb interaction between the ions contribute to an enhanced spatial confinement.
In general, each ion's excursion $z_i\,(i = 1,\cdots N)$ from the equilibrium position in a string of $N$ ions can be expressed as a linear combination of normal mode amplitudes $Z_i$ \cite{James1998}, i.e. $z_i = \sum_{j = 1}^N U_{ij} Z_j$ where $U_{ij}$ is a unitary matrix.  
Therefore, the spread $\Delta z_i$ due to thermal motion is given by
 \begin{equation}
\Delta z_{i} = \sqrt{\sum_{j=1}^{N}(U_{ij}\Delta Z_{j})^2}, \quad i = 1,\cdots N.  \label{eq:normalmodes}
\end{equation}

For two ions, this becomes $\Delta z_{1, 2}=\sqrt{((\Delta Z_{\text{COM}})^{2}+(\Delta Z_{\text{STR}})^{2})/2}$ where $\Delta Z_{\text{COM}}$ and $\Delta Z_{\text{STR}}$ are the amplitude fluctuations of the COM and stretch (STR) mode respectively. Both $\Delta Z_{\text{COM}}$ and $\Delta Z_{\text{STR}}$ obey \eqnref(\ref{eq:eignf}) and since the stretch mode has a higher eigenfrequency than the COM mode ($\omega_{\text{STR}} = \sqrt{3}\,\omega_{\text{COM}}$) they satisfy $\Delta Z_{\text{STR}} < \Delta Z_{\text{COM}}$  if both modes are at the same temperature. This results in $\Delta z_{1,2} < \Delta Z_{\text{COM}}$ which in turn means that $\Delta z_{1,2}$ is smaller than $\Delta z$ for a single ion. This argument can be extended to $N$ ions and it can be shown that $\Delta z_i < \Delta Z_{\mathrm{COM}}\,(i = 1,\cdots N)$ is satisfied in general (See Methods). 
Therefore, it is possible to achieve a higher degree of ion-localization and hence higher visibility values for a larger number of ions if the temperatures of the high-order modes are comparable to the COM mode temperature.

\figref\ref{fig:3-4-5-coupling} shows visibility measurements with three, four and five ions performed in the same frequency range as \figref\ref{fig:multi-ion-coupling}. 
Our theoretical model (shown in blue curves) fits to all the data points very well.
These results demonstrate that despite more complex variations in $V$ it is possible to achieve simultaneous cavity coupling of three to five ions  with visibility of up to 40\%. A closer look at individual data points reveals more details about the ions' geometrical configuration.
As is the case with two ions, the major local maxima in the three ions' visibility plot correspond to the ion spacing $d_1$ being an integer multiple of $\lambda/2$. Namely $d_1 =$ 20, 19 and 18 in units of $\lambda/2$ at 411, 444 and 482 kHz respectively.
However, for more than three ions, the inter-ion spacing is no longer uniform \cite{James1998}.
In the cases of four and five ions, there are two unequal distances $d_1$ and $d_2$ between the ions as shown in \figref~\ref{fig:3-4-5-coupling}. Nevertheless these inter-ion distances can be optimised to yield the best possible cavity coupling in the given frequency range. At the data points denoted as (i) and (ii) in \figref~\ref{fig:3-4-5-coupling}b and c, the following parameters are extracted from the theoretical fits: (i) $d_1 = 16.1$, $d_2 = 14.9$ and (ii) $d_1 = 16.9$, $d_2 = 15.1$  in units of $\lambda/2$.
In order to evaluate the inhomogeneities amongst the ions' cavity coupling strengths at these points, we define the normalised average coupling strength as $\tilde{g} = 1/(g_0N) \sum_{i=1}^N \abs{g(z_i^0)}$ where $g(z_i^0)$ is given by \eqnref(\ref{eq:gz}) evaluated at each ion's equilibrium position $z_i^0$ and $N$ is the total number of ions. At optimised cavity displacements  we obtain $\tilde{g} = 0.988$ and $0.983$ for the points (i) and (ii) respectively from the extracted $d_1$ and $d_2$. These results show that albeit the unequal inter-ion spacing with four and five ions we are able to couple all the ions to the cavity mode with nearly maximal strength with errors of less than 2\%.    


%

 \section{Discussion}
The work has explored a previously uncharted territory of coupling up to five trapped ions to a single cavity mode in a fully deterministic way. We have thoroughly characterised the collective coupling of multiple ions through visibility measurements of the detected cavity emission. In good agreement with the theoretical model, the results confirm our capability to control the geometrical configuration of ions to optimise the cavity coupling. An enhanced localization due to the mutual Coulomb interaction between multiple ions is observed. 
 These results make our axial-trap-cavity design an extremely attractive candidate to implement a few-qubit CQED system for studies in many-body entanglement and applications in optical quantum information processing networks.

\section{Acknowledgement}

We gratefully acknowledge support from the EPSRC (Grant No. EP/M013243/1 and EP/J003670/1).

\bibliographystyle{apsrev4-1}
\bibliography{References}

\section{Author Contributions}
S.B., M.V., H.T. and M.K. conceived the experiment. S.B., M.V. and G.K.G performed the experiment and analysed the data together with H.T. and M.K. G.K.G, H.T. and M.K. wrote the manuscript.   

\section{Additional Information}
The authors declare no competing financial interests.

\section{Methods}
\subsection{Theoretical fits for the visibility measurements}

Here we describe the theoretical model and procedure used to  produce the fitted curves in \figref\ref{fig:multi-ion-coupling} and \figref\ref{fig:3-4-5-coupling}.
In the case of multiple ions, the observed count rate of the cavity emission is proportional to
\begin{align}
 W(\nu, T, \varphi) = \sum_{i=1}^N g_{0}^{2}\left(1+e^{-k^{2}(\Delta z_i(\nu, T))^{2}}\cos(2kz_i^0(\nu)+\varphi)\right).
\end{align}
Here $z_i^0(\nu)$ and $\Delta z_i(\nu, T)$ are the equilibrium position and thermal positional spread of each ion for a given COM secular frequency $\nu$ and temperature $T$ respectively, $\varphi$ represents the overall phase offset between the cavity field and the position of the ion string and $N$ is the total number of ions. $z_i^0(\nu)$ is readily obtainable by solving coupled algebraic equations \cite{James1998} and  $\Delta z_i(\nu, T)$ can be calculated through \eqnref(\ref{eq:normalmodes}) by assuming a constant temperature for all the normal modes. For a given $\nu$ and $T$, we numerically obtain the theoretical visibility from the maximum and minimum values of $W$ by varying the offset phase $\varphi$. The fitting routine repeats this calculation for a range of the COM secular frequencies while optimising the temperature $T$ and horizontal offset $\nu_0$ as fitting parameters. The fitted $\nu_0$ values for 1-5 ion data sets are all small and within a range of 0.1-1.3 kHz.     

\subsection{Enhanced localisation in a multi-ion string}

As noted in the main text each ion's axial coordinate from the equilibrium position is given as a linear combination of normal mode amplitudes:
\begin{align}
 z_i &= \sum_{j=1}^N U_{ij} Z_{j}
\end{align}
where $N$ is the total number of ions. The matirx $U$ is made of a set of real-valued orthonormal column vectors $\{\vec{u}_j\}$ which diagonalise the harmonic expansion of the external potential and Coulomb interaction between the ions, i.e. $U = [\vec{u}_1 \vec{u}_2 \dots \vec{u}_N]$ \cite{James1998}.
The normal mode amplitudes $Z_i$ are numbered such that their eigenfrequencies are in increasing order.  Thus $Z_1 = Z_{\mathrm{COM}}$. Assuming a constant temperature for all the normal modes, $\Delta Z_i$ satisfies
 \begin{align}
   \Delta Z_i &< \Delta Z_1, \quad (i \neq 1)
 \end{align}
due to (\ref{eq:eignf}). From (\ref{eq:normalmodes}),
\begin{align}
 \Delta z_{i} &= \sqrt{\sum_{j=1}^{N}(U_{ij}\Delta Z_{j})^2}
 < \Delta Z_{1}\sqrt{\sum_{j=1}^{N}U_{ij}^2} \\
\end{align} 
From the orthonomality of $\{\vec{u}_j\}$, $\sum_{j=1}^{N}U_{ij}^2 = \sum_{j=1}^{N}(\vec{u_j})_i^2=1$. Therefore $\Delta z_{i} < \Delta Z_{1} = \Delta Z_{\mathrm{COM}}$.




\end{document}